\documentclass[%
 prc,
superscriptaddress,
 amsmath,amssymb,
 aps,twocolumn
]{revtex4-2}
\usepackage{amstext}
\usepackage{bm}
\usepackage{booktabs}
\usepackage{multirow, mathtools}
\usepackage{lineno,hyperref}
\usepackage{color}
\usepackage{flafter}
\usepackage{tabularx}
\usepackage[hang]{subfigure}
\usepackage{cancel}
\usepackage{pgfplots}
\pgfplotsset{compat=newest}
\usetikzlibrary{plotmarks}
\usetikzlibrary{arrows.meta}
\usepgfplotslibrary{patchplots}
\usepackage{grffile}
\newlength\fwidth

\newcommand{\ket}[1]{\left| #1 \right>}
\newcommand{\mv}[1]{\left< #1 \right>}

\newcommand{\vk}{v(\mfk)}
\newcommand{\uk}{u(\mfk)}
\newcommand{\up}{\uparrow}
\newcommand{\dn}{\downarrow}
\newcommand{\mf}[1]{\mathbf{#1}}
\newcommand{\+}{\dagger}

\newcommand{\mfk}{\mf{k}}

\newcommand{\ko}{k^0}
\newcommand{\nmlo}[1]{$\textrm{N}^{#1}\textrm{LO}$}
\allowdisplaybreaks

\graphicspath{{figures/}}

\begin{document}

 \title{A novel way of recasting the Bardeen-Cooper-Schrieffer gap equations}

\author{Georgios Palkanoglou}
\affiliation{TRIUMF, 4004 Wesbrook Mall, Vancouver, BC V6T 2A3, Canada}
\affiliation{Department of Physics, University of Guelph, 
Guelph, ON N1G 2W1, Canada}
\author{Alexandros Gezerlis}
\affiliation{Department of Physics, University of Guelph, 
Guelph, ON N1G 2W1, Canada}

\begin{abstract}
    The gap equations lie at the core of the Bardeen-Cooper-Schrieffer (BCS) theory, a standard tool in the description of superfluidity. As a set of non-linear integral equations, the gap equations' inherent difficulties oftentimes hinder even the crudest descriptions of superfluid states. Hard-core potentials, high-density superfluids, and coupled-channel pairing are all reasons that have historically required one to provide special treatment to the gap equations to get a solution. In this paper we present a new method for solving the gap equations that holds the promise of being an efficient universal solver that requires the minimum amount of \textit{a priori} knowledge of the targeted solutions. With theoretical evidence of exotic nuclear superfluidity posing new questions to our understanding of this fundamental property of nuclear systems, the presented method can be a valuable tool when exploring new pairing states, finite-temperature properties, or the development of sophisticated descriptions of nuclear superfludity.
\end{abstract}

\maketitle 

\section{Introduction}
Superfluids, superconductors and their properties  remain a topic of active research in many fields of physics. Nuclear superfluids and superconductors have been studied extensively due to their unique properties and the extreme conditions of their formation~\cite{dean:2003,sedrakian:2019}. Still, many properties of nuclear superfluidity remain unknown along with their effect on the description of systems of many scales, from neutron stars to nuclei~\cite{dean:2003,sedrakian:2019,allard:2025,krotscheck:2023,ding:2016,burgio:2021,fujimoto:2024,kumamoto:2024,allard:2025}.

The modeling of the microscopic mechanism underlying superfluidity often relies on a mean-field description which can either predict and postdict experiments and observations, or is the essential first step for a more sophisticated treatment of the many-body correlations~\cite{broglia:book,annett:book,tichai:2018,drissi:2024}. The first category contains many terrestrial superconductors which are dilute mixtures of weakly interacting particles and their properties can be captured by a mean-field treatment, the so-called weak coupling limit~\cite{annett:book}. On the contrary, nuclear superfluids and superconductors are prime examples of the second category, being famously strongly coupled systems, even at relatively low densities due to the strong attraction of the nuclear force. For such systems, mean-field approaches serve as a qualitative description and a guide beyond mean-field methods which are essential for a complete description~\cite{gezerlis:2010,tichai:2018,gandolfi:2022,drissi:2024,almirante:2025}. In either case, capturing the correct degrees of freedom at the mean-field level is vital to the microscopic description of superfluidity and superconductivity.

At the mean-field level, superfluidity and superconductivity are described by the BCS theory which is characterized by the gap-equations. These are a set of non-linear integral equations and they will be introduced in sec.~\ref{sec:bcs} within an overview of the BCS theory. The solution of the gap equations defines the wavefunction of the many-body system but it can be hard to identify due to their non-linearity. An additional source of complication is the two--particle interaction which is an input to these equations (see sec.~\ref{sec:bcs}). The combined problem is a famously difficult one and over the years various approaches to identifying solutions to the gap equations have been developed, tailored to intricacies of the inter-particle potential, the amplitude of the superfluid's order parameter, and in general to any \textit{a priori} knowledge of the solution to be identified~\cite{krotscheck:1972,khodel:1998,khodel:2001,ramanan:2007}. Unconventional pairing mechanisms insert evermore complicated non-linearities to the gap-equations bringing available approaches to their limit of effectiveness which in turn casts doubt on the quality of the solutions they provide: the gap-equations can have many solutions each pointing to a plausible superfluid state of the system at hand and a flexible approach that can identify all solutions is essential for the confident description of system's superfluid state~\cite{goodman:1972,takatsuka:1993,goodman:2001,bulgac:2006,gezerlis:2011,ding:2016,palkanoglou:2024,fujimoto:2024,guo:2025}.

{\nobreak In this paper we develop a novel approach to solving the gap equations that is agnostic to the specifics of the physical system. It can be straightforwardly extended to any form of gap-equation that satisfies some general requirement. Our proposed approach holds thus the promise of a universal solver of the gap equation that can be used to build descriptions of any type of superfluid at the mean-field level. The rest of this paper is organized as follows: in sec.~\ref{sec:bcs} we provide an overview of the BCS theory and introduce the gap equations therein along with the traditional approach to their solution; for more sophisticated approaches the reader is referred to elsewhere. In sec.~\ref{sec:recast} we introduce a novel approach to solving the gap equations that is not tailored to a specific physical system and in sec.~\ref{sec:calculations} we put it to use in the case of superfluid neutron matter, solving the gap equations for a range of neutron-neutron potentials, comparing to existing approaches, and prescribing how to avoid edge-cases. Finally, sec.~\ref{sec:conclusions} is dedicated to the discussion of the results and the outlook of the new method for solving the gap equation.

\section{An Overview of standard BCS theory}
\label{sec:bcs}
Before presenting a recasting of BCS theory's cornerstone, the gap equations, we will present an overview of BCS theory and the origin of the gap equations. For a more in-depth discussion of the BCS theory and its use in describing nuclear superfluids see, for example, Refs.~\cite{annett:book,takatsuka:1993,dean:2003,sedrakian:2019} and references therein. Here, and throughout the paper we will focus on s-wave BCS theory, that is, the mean-field theory of superfluids composed of pairs with relative angular momentum $l=0$. We chose this to simplify the discussion since the proposed recasting can be straightforwardly extended to other $l$-channels, coupled or uncoupled.  We will come back to this topic in sec.~\ref{sec:conclusions}. }

The BCS theory assumes the form of the ground state of a superfluid system to be
\begin{align}
\ket{\psi} = \prod_\mfk \left[\uk + \vk  c_{\mfk_\up}^\+c_{-\mfk_\dn}^\+\right]\ket{0} ~,\label{eq:state}
\end{align}
which describes a condensation of pairs: each $\mfk$-level is occupied by a pair with a probability $v^2(\mfk)$. Here, $c^\+_{\mfk\sigma}$ is the creation operator for a fermion with momentum $\mfk$ and spin projection $\sigma$, and they are associated with solutions of the single-particle Schr{\"o}ndiger's equation. As we are often interested in bulk properties of a superfluid, these single-particle states are normalized in a box of length $L$, under Periodic Boundary Conditions (PBC). This is to guarantee translational invariance, and the thermodynamic limit (TL), i.e., $L\to \infty$ should be eventually  taken. In the standard BCS approach, all normal-state interaction are neglected in the Hamiltonian which includes only the pairing interaction and takes the form
\begin{align}
H= \sum_{\mfk} \epsilon_\mfk c_{\mfk\sigma}^\+ c_{\mfk\sigma} + \sum_{\mfk\mfk'}V_0(\mfk,\mfk')c_{\mfk\up}^\+ c_{-\mfk\dn}^\+  c_{-\mfk\dn} c_{\mfk\up} \label{eq:ham}
\end{align}
where $\epsilon_\mfk=\hbar^2\mfk^2/2m$ is the free single particle spectrum and only pairs with no centre-of-mass momentum are taken into account. For nuclear condensates, like the neutron superfluids or the proton superconductors found in neutron star matter, the potential in Eq.~(\ref{eq:ham}) is expanded in angular momentum channels and the one dominating for the relevant densities is kept. For neutrons, which will be our topic, that is the $l=0$ channel and $V_0(k,k')$ stands for the projection of a potential $V(r)$ on that channel:
\begin{align}
    V_0(k,k') = \int_0^\infty dr r^2 j_0(kr)V(r)j_0(k'r)~, \label{eq:pot}
\end{align}
where $j_0(x)$ is the zero-th order spherical Bessel function of the first kind. Note that even though Eq.~(\ref{eq:pot}) implies that $V_0(k,k')$ is defined via a coordinate space potential $V(r)$, that is not necessary and a two-particle potential can also be defined directly in momentum space. In sec.~\ref{sec:calculations} we will employ one such potential. From Eq.~(\ref{eq:ham}), the free energy of the state is
\pagebreak[4]
\begin{align}
    W&(\{\vk\};\mu) = \mv{H}-\mu \mv{N} \notag\\
    &= \sum_{\mfk\sigma} \xi_\mfk \vk^2 + \sum_{\mfk\mfk'}V(\mfk,\mfk')\uk\vk u_{\mfk'}v_{\mfk'}~,
\end{align}
where $\xi_\mfk=\epsilon_\mfk-\mu$ is the single-particle excitation spectrum shifted by the chemical potential $\mu$. The minimum of the BCS state's free energy corresponds to the distribution $\vk$ that satisfies the gap equation
\begin{align}
\Delta(\mfk) = -\frac{1}{2}\sum_{\mfk'}V(\mfk,\mfk') \frac{\Delta(\mfk)}{E(\mfk)} \label{eq:gap_vec}
\end{align}
where
\begin{align}
    v^2(\mfk) = \frac{1}{2}\left(1-\frac{\xi_\mfk}{E_\mfk}\right)
\end{align}
where $E_\mfk = \sqrt{\xi_\mfk^2+\Delta_\mfk^2}$ is the quasiparticle excitation spectrum. The gap function $\Delta(\mfk)$ describes the binding energy of a pair of particles with momentum $\mfk$ and it uniquely defines the superfluid state in Eq.~(\ref{eq:state}). The energy of the condensate is then a functional of $\Delta(\mfk)$:
\begin{align}
    \mv{H}\left[\Delta\right] = \sum_{\mfk}\left[\epsilon_\mfk\left(1-\frac{\xi_\mfk}{E_\mfk}\right) - \frac{\Delta^2(\mfk)}{E(\mfk)}\right] ~.\label{eq:energy}
\end{align}
The gap equation is often complemented by the average particle-number equation:
\begin{align}
    \mv{N} = \sum_\mfk 2v^2(\mfk) = \sum_\mfk \left(1-\frac{\xi_\mfk}{E_\mfk}\right) ~. \label{eq:avn}
\end{align}
Note that the trivial gap function, $\Delta(\mfk)=0$ for all $\mfk$, is a solution of the gap equation for any interaction $V$. This is essential in concluding that the ground-state of a given system is a superfluid described by Eq.~(\ref{eq:state}): for a given interaction $V(\mfk,\mfk')$, a second non-trivial solution $\Delta'(k)$ will always correspond to a lower energy: $\mv{H}[0]>\mv{H}[\Delta']$. Hence, the existence of a second solution of Eq.~(\ref{eq:gap_vec}) is generally considered a definition of pairing correlations~\cite{frauendorf:2014}.

At the thermodynamic limit, i.e., $L\to \infty$, $\mv{N}\to \infty$, and $\mv{N}/L^3 \to \rho$, the sums in Eqs.~(\ref{eq:gap_vec}) and (\ref{eq:avn}) turn into integrals and Eq.~(\ref{eq:avn}) determines the superfluid density:
\begin{align}
\Delta(k) &= -\frac{1}{\pi}\int_0^\infty dk' k'^2 V_0(k,k') \frac{\Delta(k')}{E(k')} \label{eq:gap}\\
 \rho &= \frac{1}{2\pi^2} \int dk k^2 \left(1-\frac{\xi(k)}{E(k)}\right)\label{eq:abn.int}
\end{align}
where $V_0(k,k')$ is given by Eq.~(\ref{eq:pot}). The gap equations in other channels of $l$ are similar, in the absence of channel mixing. As already mentioned, we will focus on the s-wave gap equation, as it is the most relevant one for neutron matter, but all the presented arguments can be straightforwardly modified for other partial-wave channels.

Though not typically cast in this mathematical language, Eq. (\ref{eq:gap}) is a nonlinear  integral equation of the Hammerstein type~\cite{atkinson:1992}:
\begin{align}
    \Delta (k) = \mathcal{K}_k\left[\Delta(k)\right] = \int dk' K(k,k') f\left(k',\Delta(k')\right)
\end{align}
where square brackets signify functional dependence and we identify
\begin{align}
    K(k,k') &= -\frac{1}{\pi} k'^2 V_0(k,k') \\
    f(k',\Delta(k')) &= \frac{\Delta(k')}{\sqrt{\xi^2(k') + \Delta^2(k')}}
\end{align}
 In current standard practice, an iterative approach is employed to solve this equation where assuming a trial distribution $\Delta^{(0)}(k)$, convergence is reached through a sequence of distributions $\Delta^{(n)}(k)$. The integrals are evaluated numerically with a quadrature method turning them into sums of the functions involved on a grid. This way, the distribution $\Delta^n(k)$ is represented on a $k$-grid of length $m$ by a $m$-dimensional vector $\Delta_i^{(n)}=\Delta^{(n)}(k_i)$, $i=1,\dots,m$. The solution via iteration defines a sequence of vectors $\Delta^{(n)}=\{\Delta^{(n)}_i\}$, $n=1,2,\dots$, via the action of the Hammerstein operator $\mathcal{K}$, 
 \begin{align}
     \Delta_i^{(n)} &=\mathcal{K}_{k_i}\left[\Delta^{(n-1)}\right]= \mathcal{K}_{k_i}^n\left[\Delta^{(0)}\right]
 \end{align}
 where we also defined the $n$-th power of the Hammerstein operator as $n$ applications of the operator as $\mathcal{K}_k^n[\cdot] = \mathcal{K}_k \left[\mathcal{K}_k\left[\cdots \mathcal{K}_k\left[\cdot \right]\right]\right]$. In practice, this is done iteratively, starting from a trial vector $\Delta^{(0)}$, defining the vector $\Delta^{(1)}=\{\Delta_i^{(1)}\}$ with $\Delta_i^{(1)}=\mathcal{K}_k[\Delta^{(0)}]$, and so on. Convergence is determined when the relative difference of the successive vectors,
 \begin{align}
     \delta \Delta^{(n)} = \sum_i\left|\frac{\Delta_i^{(n-1)}-\Delta_i^{(n)}}{\Delta_i^{(n-1)}}\right|~,
 \end{align}  
 is lower than some small number. This method is often called the direct iteration method. A modified version of the direct iteration method has been developed in which at step $n$ the Hammerstein operator is applied on a linear combination of the vectors from steps $n-1$ and $n-2$ which are mixed with a learning rate that is tuned to the specifics of the system. This learning rate often requires fine tuning~\cite{drischler:2017}. A more sophisticated iterative method based on Broyden's method for solving general non-linear equations, named modified Broyden's method, has also been developed where one operates similarly to the direct iteration method but amplifies the vector $\Delta_i^{(n-1)}$ that the Hammerstein operator is acting on with even more information~\cite{johnson:1988}. Finally,  Ref.~\cite{khodel:2001} introduced methods that decouple the total scale of the vector $\Delta_i$, parametrized by its value at the Fermi surface $\Delta_F=\Delta_{k_i=k_F}$. These are especially effective when looking for solutions with small $\Delta_F$ where we are faced  with a semi-singularity. Still, interparticle potentials that vanish for momenta close to the Fermi surface are handled separately.

 In all methods mentioned above a solution is approached iteratively with a Hammerstein operator acting on a vector at each iteration. This has two major implications: the grid imposed by the quadrature method must match the grid used to discretize the distribution $\Delta(k)$ and an integral evaluation is performed at each step. These features mean that the iterative methods scale badly with the precision of the integral in the Hammerstein operator and high precision is often needed for convergence as gap functions are often a fraction of the system's total energy. Additionally, simple iteration methods grant little freedom in the path that leads to convergence, often making finding non-trivial solutions difficult. The method presented in this paper holds the promise of ameliorating these issues.
 
\section{A new collocation-type method for the recast gap equations}
\label{sec:recast}
The available methods for solving the gap equations suffer from various peculiarities discussed in the previous sections. In this section we introduce a method which is based on collocation inspired by ideas proposed in Refs.~\cite{kumar:1987a} and \cite{kumar:1987b} and, in short, it solves for the kernel of the gap equations, instead of the gap function itself, on a collocation grid. We claim that this method can alleviate many issues plaguing the iteration-based approaches mentioned above.

Before introducing the collocation method for the gap equation's kernel, let us briefly discuss collocation methods for the gap function which, even though rarely discussed, will serve as an effective introduction to the new method. In a collocation method for the gap function one expands the gap function on a set of linearly independent functions recasting the gap equation into a algebraic equation for the expansion coefficients. For a set of known basis functions $\left\{u_i\right\}$ the gap function can be expanded as
\begin{align}
\Delta(k) = \sum_{i=1}^n d_{i} u_{i}(k)~.
\end{align}
 The unknown coefficients $d_{i}$ are found by requiring $\Delta$ to satisfy the gap equation, i.e., Eq.~(\ref{eq:gap}),
\begin{align}
\Delta\left(k_{i}\right) = -\frac{1}{\pi} \int_0^\infty  dp p^2 V_0(k_{i},p)\frac{\Delta\left(p\right)}{\sqrt{\xi^2\left(p\right)+\Delta^2\left(p\right)}}~,\label{eq:gapeq}
\end{align}
with $i=1,\dots,n$ enumerating the collocation grid $\left\{k_{i}\right\}$ of $n$ distinct points in $\left[0,\infty\right)$. 
This gives rise to $n$ non-linear equations for $d_{i}$,
\begin{align}
\sum_{i=1}^n & U_{ji} d_i =-\frac{1}{\pi} \sum_{i=1}^n d_i ~\times\notag 
\\
&\times~\int dkdk' \frac{u^*_j(k)V_0(k,k')u_i(k')}{\sqrt{\xi^2(k')+ \left[\sum_{i=1}^n d_i u_i(k')\right]^2}}~,\label{eq:gap.col.d}
\end{align}
with $j=1,\dots,n $ and $U_{ij}=\int dk u_i^*(k)u_j(k)=\delta_{ij}$, if the basis is orthonormal. In practice, Eq.~(\ref{eq:gap.col.d}) can be solved by some iterative scheme, similar to those employed for the standard gap equations and were described in the previous section. Such approaches would still suffer from similar issues as the solvers for the standard gap equation, for example, a strong dependence on the value of $\Delta_F$ due to a semi-singularity at the Fermi surface remains. We will attempt to ameliorate these issues with a new solver.

Inspired by Refs.~\cite{kumar:1987a,kumar:1987b} we employ a collocation applied to an equivalent equation for $F(k)=\Delta(k)/E(k)$.
In terms of $F(k)$, the gap equation is
\begin{align}
\Delta(k) &= -\frac{1}{\pi} \int dp p^2 V(k,p) F(p)~, \label{eq:gap_amp}\\
F(k)&= \frac{\Delta(k)}{\sqrt{\xi^2(k)+\Delta^2(k)}}~,\label{eq:condensation_amp}
\end{align}
and through these we can define an equation for $F(k)$, equivalent to the gap equation:
\begin{align}
F(k) =\frac{-\frac{1}{\pi} \int dp p^2 V(k,p) F(p)}{\sqrt{\xi^2(k)+\left[\frac{1}{\pi} \int dp p^2 V(k,p)F(p)\right]^2}} ~. \label{eq:recast_f}
\end{align}
Applying a collocation method on Eq.~(\ref{eq:recast_f}) amounts to approximating $F(k)$ by
\begin{align}
F_n(k) = \sum_i^n g_i \phi_i(k)~\label{eq:F-interpol}
\end{align}
on a set of basis functions $\left\{\phi_i\right\}$, with $g_i$ satisfying Eq.~(\ref{eq:recast_f}) on a collocation grid $\left\{k_j\right\}$,
\begin{align}
\sum_i^n g_i \phi_i\left(k_j\right) = \frac{-\frac{1}{\pi}\sum_i^n g_i \int dp p^2 V(k_j,p)  \phi_i(p)}{\sqrt{\xi^2(k)+\left[\frac{1}{\pi} \sum_i^n g_i \int dp p^2 V(k_j,p)\phi_i(p)\right]^2}}~,\label{eq:recast.col}
\end{align}
with $j=1,\dots,n$. Crucially, the integrals in Eq.~(\ref{eq:recast.col}) do not depend on the set of $\left\{g_i\right\}$ and so they only need to be evaluated once. The non-analytic nature of the square-root for small arguments in the denominator of Eq.~(\ref{eq:recast.col}) creates problems similar to the semi-singularity of the standard gap equations. We can alleviate this ill-conditioning by rearranging and squaring Eq.~(\ref{eq:recast.col}), finally leading to
\begin{align}
\left(\sum_i g_i\phi_{ij}\right)^2&\left[\xi^2(k_j)+\left(\sum_{i}g_i\psi_{ij}\right)^2\right] - \notag \\
&-\left(\sum_i g_i\psi_{ij}\right)^2 = 0 \label{eq:recast_raw}
\end{align}
where we defined
\begin{align}
\psi_{ij} &=\psi_i(k_j) =  -\frac{1}{\pi}\int dp p^2 V(k_j,p)  \phi_i(p) \label{eq:psi_raw}\\
\phi_{ij} &= \phi_i(k_j) \label{eq:phi_raw}
\end{align}
Equation~(\ref{eq:recast_raw}) is a non-linear equation, in terms of $n$ unknowns $g_i$ and can be solved by a multi-dimensional root-finding method. It is a recast gap equation and, as we will show in the next few sections, it is free of some major issues of the standard gap equations. 

Before carrying on with solving the recast gap equation in Eq.~(\ref{eq:recast_raw}), a few comments on its equivalence to the standard gap equations are in order. The one-to-one correspondence between the solutions of Eqs.~(\ref{eq:gap}) and (\ref{eq:recast_raw}) can be proven if~\cite{kumar:1987a}
\begin{align}
\sup_{0\le k <\infty} \int_0^\infty dp p^2 \left|V_0(k,p) \right| & < \infty \label{eq:cond1}\\
\lim_{k\to k'}\int_0^\infty dpp^2 \left|V_0(k,p)-V_0(k',p)\right|&=0 ~, \notag \\
\textrm{for all}~k'\in\left[0,\infty\right)&\label{eq:cond2} \\
g(k,v) = \frac{v}{\sqrt{\xi^2(k)+v^2}} ~&: \notag\\
\quad\textrm{continuous in}~\left[0,\infty\right)\times &\mathbb{R}~. \label{eq:cond3}
\end{align}
The first condition holds for all relevant potentials because $V_0(k,p)$ can be seen as the scattering amplitude in the Born-approximation. The second condition is satisfied for all continuous potentials, e.g., the kind in Eq.~(\ref{eq:pot}) which inherit their continuity from the spherical Bessel functions. The last condition is true for any continuous dispersion relations $\epsilon(k)$. These conditions are met for most realistic descriptions of nuclear systems rendering Eq.~(\ref{eq:recast_raw}) equivalent to Eq.~(\ref{eq:gap}) for nuclear superfluids.

\begin{figure}[htp]
\centering
\includegraphics[width=0.45\textwidth]{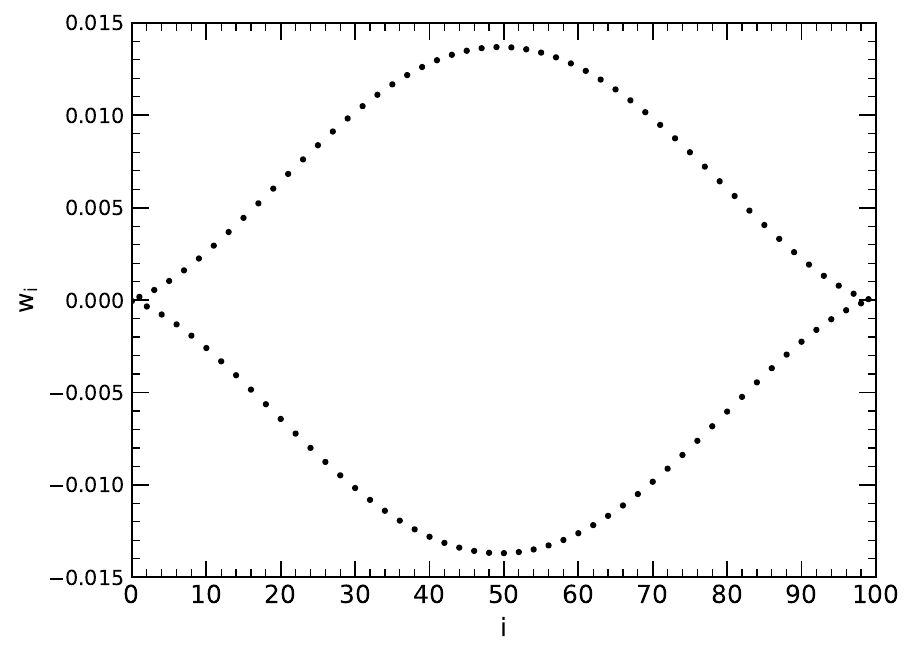}
\caption{The weights for the barycentric formula in Eq.~(\ref{eq:bary}) for 100 Legendre nodes in the interval $[0,10]$.}
\label{fig:leg_nodes}
\end{figure}

The recasting of Eq.~(\ref{eq:recast_raw}) already provides advantages which we will elucidate below, but as it stands it is not a well-defined problem without a choice for the basis functions $\{\phi_i\}$ and the grid $\{k_i\}$. More importantly, the right choices for these elements could make Eq.~(\ref{eq:recast_raw}) a highly effective recasting of the gap equation.

When choosing the basis functions $\{\phi_i\}$ in Eq.~(\ref{eq:phi_raw}) the only formal requirement is that they must be a set of linearly independent functions. When the stronger condition of orthogonality is not enforced, $F(k_i)= \sum_{j=1}^n g_j\phi_j(k_i)$ represents an interpolation of $F(k)$ on the grid $\{k_i\}$ . The quality of the interpolation and of the collocation depends on the choice of $\{\phi_i\}$ (and $\{k_i\}$, as discussed below) and an efficient choice is cardinal polynomials anchored on a grid $\{k_i\}$,
\begin{align}
\phi_i(k) &= \frac{w_i}{k-k_i} \left[\sum_j\frac{w_j}{k-k_j}\right]^{-1}~. \label{eq:bary}
\end{align}
Here we have written the cardinal polynomials using the barycentric formula~\cite{berrut:2004,gezerlis:book} with the barycentric weights,
\begin{align}
    w_i = \frac{1}{\prod_{j\neq i}(k_i-k_j)}~. \label{eq:weights}
\end{align}
which depend entirely on the anchoring grid and so they carry much information about it. Most importantly, the quality of the anchoring grid can be assessed by the dispersion of the weights $w_i$'s because weights that vary widely would make $F(k)$ sensitive to small variations of the $g_i$'s and they would signify a non-optimal choice of anchoring grid. In that regard, anchoring grids with high point-density at the ends of the interpolation interval mitigate that problem, with typical examples being Chebyshev nodes or Legendre nodes, the zeros and extrema of the Chebyshev or Legendre polynomials, respectively~\cite{berrut:2004}.


\begin{figure}[htp]
\centering
\includegraphics[width=0.45\textwidth]{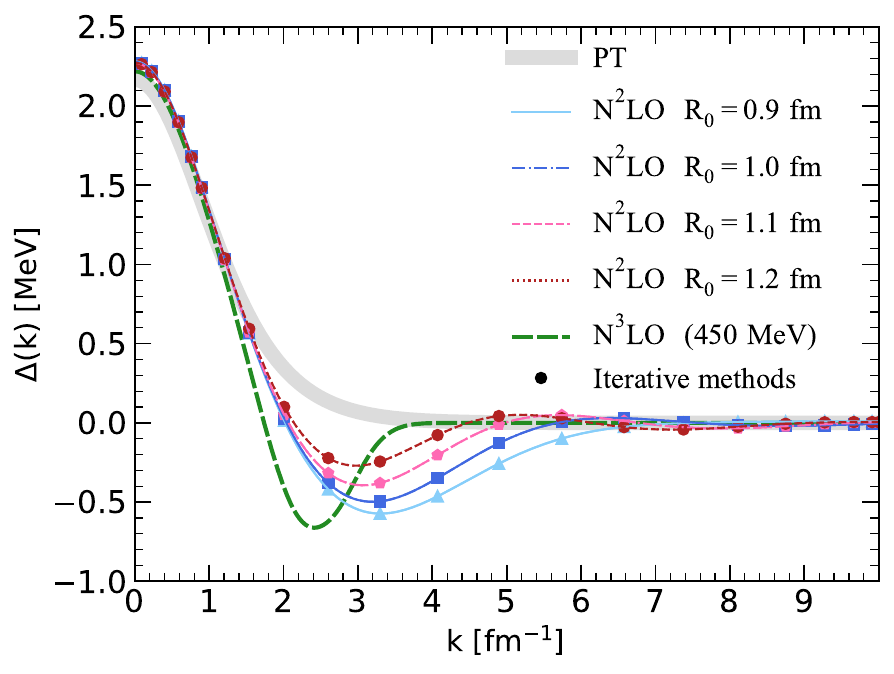}
\caption{Solutions of the gap equations with various potentials: P{\"o}schl-Teller potential (grey solid line), \nmlo{2} potential with $R_0=0.9~\textrm{fm},~1.0~\textrm{fm},~1.1~\textrm{fm},~1.2~\textrm{fm}$ (solid light blue, dashed-dotted navy blue, dashed red, dotted dark red), and \nmlo{3} (450 MeV) potential (green dashed). Solutions from iterative methods are included for the \nmlo{2} potential (solid circles).}
\label{fig:gaps}
\end{figure}

Sets of weights that don't vary widely, but are of very large or very small overall scale could lead to over- or underflow, but that can also be avoided because the barycentric formula in Eq.~(\ref{eq:bary}) does not depend on the overall scale of the weights. Indeed, for a large number of points, $n_k$, in an interval $[k_0,k_1]$, the scale of the weights in Eq.~(\ref{eq:weights}) scales as $C^{-n_k}$, where $C=(k_0-k_1)/4$ is called the capacity of the interval $[k_0,k_1]$. Then if all factors of $(k_i-k_j)$ in Eq.~(\ref{eq:weights}) are multiplied by $C^{-1}$, the resulting weights are rescaled removing this asymptotic behavior making the case of over- or underflow unlikely. In general, any linear transformation of the interval leaves the weights unchanged (this additionally signifies that a linear transformation of the grid cannot improve its quality for interpolation). To demonstrate all of the above, we plot the rescaled weights corresponding to 100 Legendre nodes in the interval $[0,10]$ in Fig.~\ref{fig:leg_nodes}. These are identical to the weights corresponding to 100 Legendre nodes in the interval $[-1,1]$ and the small amplitude of their variation means a well-chosen anchoring grid: small variations of the $g_i$'s in Eq.~(\ref{eq:F-interpol}) will not be catastrophic.

Finally, we discuss the choice of the anchoring grid $\{k_i\}$ which must be considered in tandem with two other grids: the discretization grid, which is the grid enumerated by the index $j$ in Eq.~(\ref{eq:recast_raw}), and the quadrature grid used to evaluate the integral in Eq.~(\ref{eq:psi_raw}). In short, while there is no formal requirement for these grids to match, it is most efficient to use the roots of Legendre polynomials for all three purposes. In detail, anchoring of the cardinal polynomials on a Legendre grid greatly facilitates the interpolation because of its high point-density at the end of the interval~\cite{berrut:2004}. Then, choosing the same grid for discritization eliminates a sum in Eq.~(\ref{eq:recast_raw}) because it means that $\phi_{i}(k_j)=\delta_{ij}$. Finally, choosing a Legendre grid for the quadrature of the integral in Eq.~(\ref{eq:psi_raw}), i.e., performing a Gauss-Legendre quadrature, eliminates the integral all together making $\psi_{ij}$ proportional to the matrix element of the potential. Note that for further efficiency, we use a grid of length $n\times N$ made of $n$ Legendre grids of order $N$ stitched together (see sec.~\ref{sec:calculations} for details). The downside of choosing the same grid for collocation and for quadrature is that each $\psi_{ij}$ contains information of only $V_0(k_i,k_j)$ meaning that capturing long-tailed potentials would require a grid that extends to high momenta. We mitigate this problem by transforming the tails of our grids with a joint, as described below.

\begin{figure}[htp]
\centering
\includegraphics[width=0.45\textwidth]{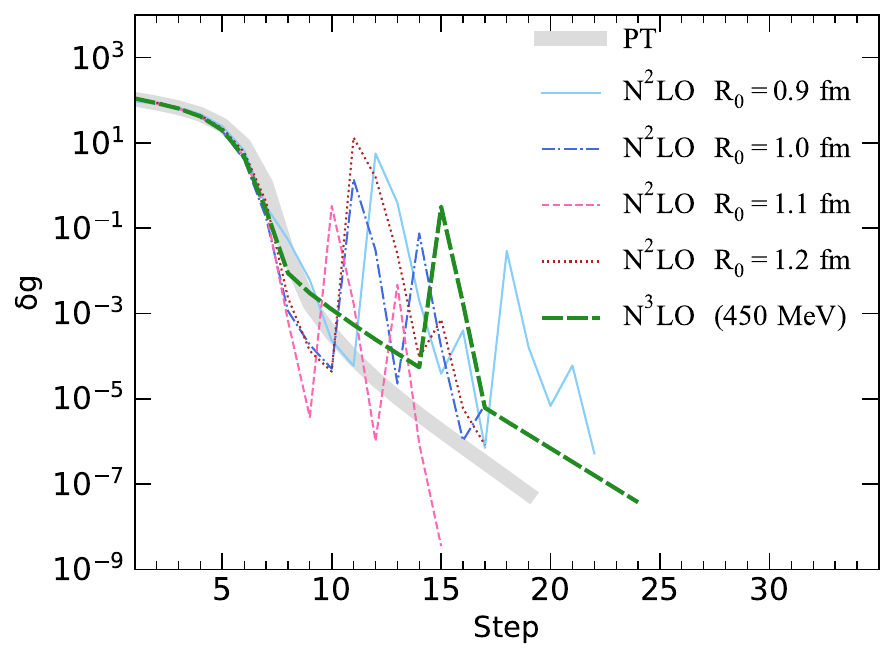}
\caption{Convergence of Newton's method in $\delta g$ for the PT potential (thick grey solid line), the \nmlo{2} potentials with $R_0=0.9~\textrm{fm}$ (blue solid line), $R_0=1.0~\textrm{fm}$ (dark blue dashed dotted line), $R_0=1.1~\textrm{fm}$ (dashed red line), $R_0=1.2~\textrm{fm}$ (dotted red line), and the \nmlo{3} potential (green dashed line). Three regimes are separated by the rate of convergence and analyzed in Fig.~\ref{fig:regimes}.}
\label{fig:conv}
\end{figure}

The number of points $N$ needed to solve Eq.~(\ref{eq:recast_raw}), is determined by the potential in $\psi_{ij}$: a long-tailed potential will generate a matrix $\psi_{ij}$ with non-zero elements farther from the diagonal. This happens when the potential contains a strong short-range repulsion in coordinate space, a typical case for nuclear physics. To make sure that the $k$-space is sufficiently covered we can stretch the tail of our grid by the use of a joint. In other words, we can map the portion of a $k$-grid that extends from some value $k_0$ to $\infty$ to a finite interval:
\begin{align}
\left[0,\infty\right] &= \left[0,k_0\right]\cup \left(k_0,\infty\right) \mapsto \notag\\
&\mapsto\left[0,k_0\right]\cup \left[\lambda(k_0),\lambda(\infty)\right] = [t(0),t(\infty)]~.
\end{align}

An efficient choice for the mapping $t$ is given in the appendix~\ref{app:joint}. With the use of a joint and discritized on the anchoring grid, Eq.~(\ref{eq:recast_raw}) takes its final form:
\begin{align}
f_i(\{g\}) = \tilde{g}_i^2 &\left[\tilde{\xi}_i^2+\left(\sum_j \tilde{\psi}_{ij}\tilde{g}_j\right)^2\right]-\notag \\
&-\left(\sum_j \tilde{\psi}_{ij}\tilde{g}_j\right)^2 = 0 ~, \label{eq:recast}
\end{align}
where
\begin{align}
\tilde{\xi}_i &=\xi\left[t^{-1}(x_i)\right]~, \\
\tilde{\psi}_{ij} &=   -\frac{1}{\pi}  c_j\left[t^{-1}(x_j)\right]' {\left[t^{-1}(x_j)\right]^2} V\left[t^{-1}(x_i),t^{-1}(x_j)\right]~,
\end{align}
and $c_j$ is the $j$-th weight associated with the Gauss-Legendre quadrature. Once again, the mapping function $t$ maps the tail of the finite $k$-grid to large $k$-values and the choice made for the applications presented in the next sections is detailed in the appendix~\ref{app:joint}.

\begin{figure}[htp]
\centering
\includegraphics[width=0.45\textwidth]{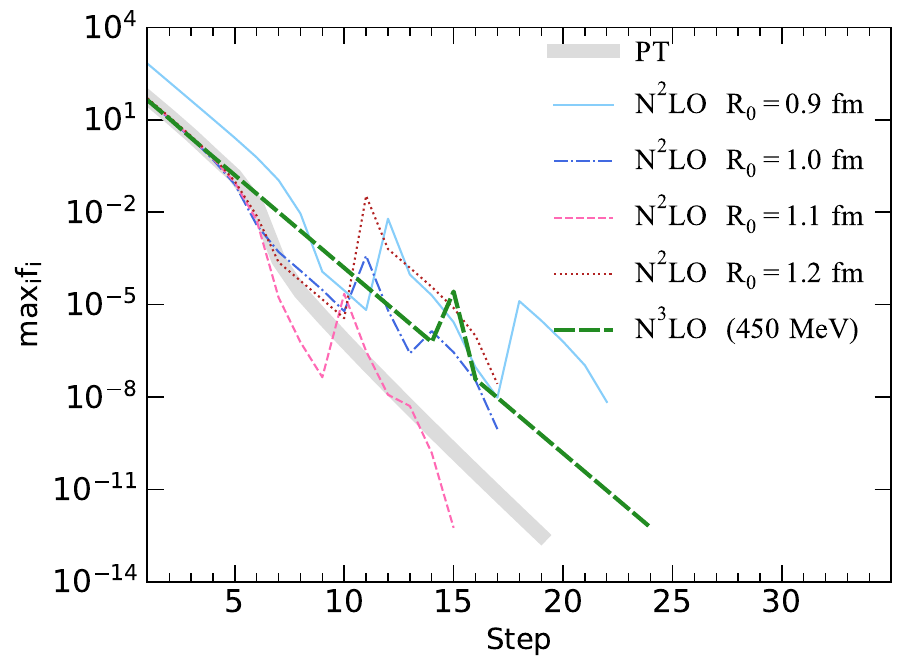}
\caption{Convergence of Newton's method in $\textrm{max}_if_i$ for the PT potential (thick grey solid line), the \nmlo{2} potentials with $R_0=0.9~\textrm{fm}$ (blue solid line), $R_0=1.0~\textrm{fm}$ (dark blue dashed dotted line), $R_0=1.1~\textrm{fm}$ (dashed red line), $R_0=1.2~\textrm{fm}$ (dotted red line), and the \nmlo{3} potential (green dashed line).}
\label{fig:maxf}
\end{figure}

The set of Eqs.~(\ref{eq:recast}) poses a well-defined problem which can be solved with any root finding algorithm. To provide a fair comparison with existing methods, we employ a standard multidimensional version of Newton's method, as described in Ref.~\cite{gezerlis:book}. Newton's method finds the solution by stepping through different distributions $g_i$ iteratively guided by the equations' Jacobian matrix $\partial f_i/ \partial g_j$ and typically converging quadratically. We find that when employing a simple solver for Eqs.~(\ref{eq:recast}), the recast method outperforms standard approaches in various ways, promising even better performance if a more sophisticated method is used. In what follows, we estimate convergence by comparing successive $g_i$ distributions with the following metric
\begin{align}
    \delta g ^{(n)} = \frac{\sum_i |g^{(n)}_i-g^{(n-1)}_i|}{\textrm{max}_i g_i^{(n)}}~.
\end{align}
while also displaying the maximum of the set of the non-linear equations in Eq.~(\ref{eq:recast}) , namely, $\textrm{max}_i f_i$ as an alternative more conservative convergence metric.

\begin{figure}[htp]
\centering
\includegraphics[width=0.5\textwidth]{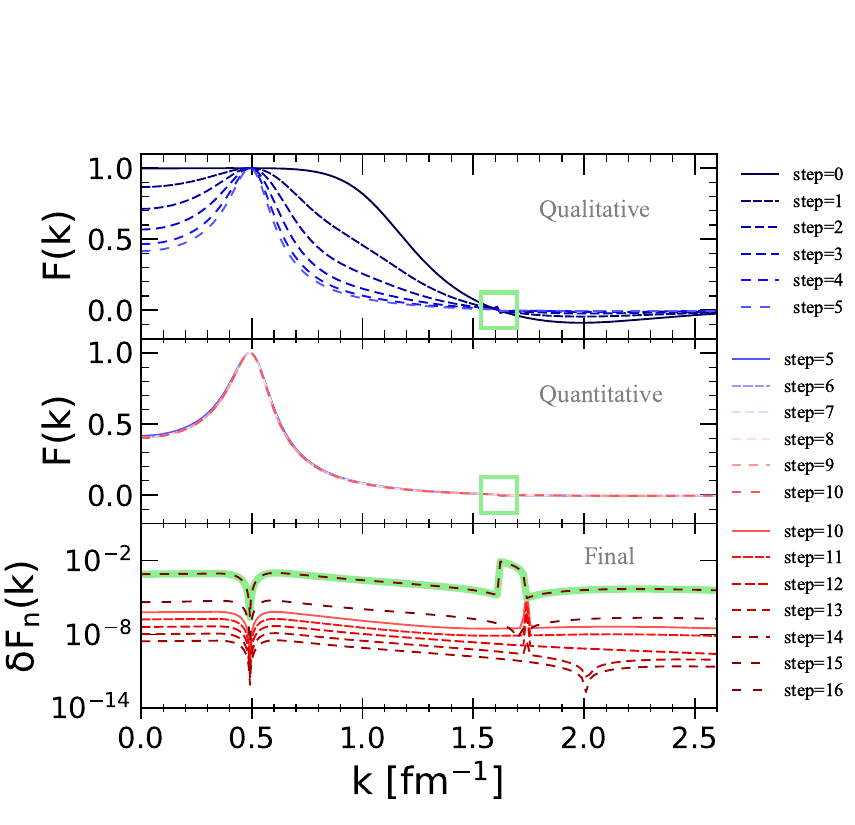}
\caption{The three convergence regimes of $F(k)$ for the \nmlo{3} potential: a qualitative regime ($0<\textrm{step}\lessapprox5$, top panel), a quantitative regime ($5<\textrm{step}\lessapprox10$, middle panel), and a final regime ($10<\textrm{step}$, bottom panel). The green squares in the top and bottom panels mark the position of discontinuity created at the node of $F(k)$ while the green curve in the bottom panel highlights the effect of correcting it in the way described in sec~\ref{sec:smooth}.}
\label{fig:regimes}
\end{figure}

\section{Calculations}
\label{sec:calculations}
We will now use the recast method to perform calculations of the pairing gap in neutron matter for a purely attractive potential and a number of realistic nuclear interactions. We will verify the accuracy of the results with the other methods mentioned above, e.g., direct iteration or Khodel's method, and compare performances. These calculations employ a grid of 1500 points made of three Legendre grids spanning the intervals $(0,1)$, $(1,10)$, and $(10,51)$. Additionally, the interval $(50,51)$ is mapped onto $(50,\infty)$, using the joint described in sec.~\ref{sec:recast} and appendix~\ref{app:joint}, with a large number is chosen instead of $\infty$; we chose $400$ for the results below. The density of this grid is rather moderate compared to the ones typically used for high-quality calculations of nuclear pairing gaps and it was chosen to highlight the recast method's performance. The same chemical potential was chosen for all calculations $\mu=5~\textrm{MeV}$ (or $k_\textrm{F}\approx 0.5~\textrm{fm}^{-1}$), unless specified otherwise.

We chose a number of nuclear potentials with different properties in coordinate space. The simplest one is the P{\"o}schl-Teller (PT) potential which is purely attractive and tuned to reproduce low-energy neutron-neutron scattering~\cite{pera:2023}. We also employed the realistic neutron-neutron interactions \nmlo{2} and \nmlo{3} from Refs.~\cite{gezerlis:2013,entem:2003}, respectively. For the \nmlo{2} potential, which is formulated in coordinate space, we explored the dependence of the new method's performance on the ``hardness'' of the interaction, that is, the amplitude of the potential's repulsive core in coordinate space. The resulting gap functions are shown in Fig.~\ref{fig:gaps}. These gap functions reproduce the solutions from existing methods and we illustrate this by comparing our new method's results for the \nmlo{2} potentials with results from iterative methods (filled circles) in the same Fig.~\ref{fig:gaps}. The major advantage of our approach is its rapid and system-independent convergence and in Figs.~\ref{fig:conv} and \ref{fig:maxf} we demonstrate this by plotting two convergence metrics: $\delta g$, i.e.,  the difference of the distribution $g_i$ from that of the previous iteration, and $\textrm{max}_i f_i$, respectively, for all iterations of the multi-dimensional Newton's method. We notice that the eventual convergence doesn't depend on the interaction chosen with all calculations converging within 30 steps. As is evident from Fig.~\ref{fig:conv}, the convergence can be separated into three regimes for all potentials: a qualitative regime ($0<\textrm{step}\lessapprox5$), a quantitative regime ($5\lessapprox\textrm{step}\lessapprox10$), and a final regime ($10\lessapprox\textrm{step}$). The evolution of $F(k)$ through each regime for the \nmlo{3} potential at $\mu=5~\textrm{MeV}$ is shown in Fig.~\ref{fig:regimes}.

The three regimes are distinguished by the step-by-step changes in $F(k)$ observed in each. In the qualitative regime ($0<\textrm{step}\lessapprox5$), $F(k)$ converges on its qualitative features, as the regime's name suggests and as seen in the top panel of Fig.~\ref{fig:regimes}; this takes $\sim5$ steps. In the quantitative regime ($5\lessapprox\textrm{step}\lessapprox10$), presented in the middle panel of Fig.~\ref{fig:regimes}, the qualitative features of $F(k)$ have converged and remain unchanged while some values of $F(k)$, especially at low-$k$ converge quantitatively. In the final regime ($10\lessapprox\textrm{step}$), changes of the order of $10^{-6}$ are observed which are plotted in the bottom panel Fig.~\ref{fig:regimes} in the form of the difference $\delta F_n(k)=|F_{n}(k)-F_{n-1}(k)|$, at each step $n>10$. At step $n=15$ we course-correct a discontinuity in $F(k)$ generated by Newton's method (see sec.~\ref{sec:smooth} for details) resulting in a generally larger $\delta F_n(k)$ highlighted with a green line in the bottom panel of Fig.~\ref{fig:regimes}. This discontinuity lies at the center of the green square in the top and middle panels of Fig.~\ref{fig:regimes}. While Fig.~\ref{fig:regimes} demonstrates the convergence of $F(k)$ for the \nmlo{3} interaction, the existence and extent of the three regimes, where $F(k)$ converges at different scales, seems to be a largely potential-independent feature that one can use to estimate the quality of the solution at each step.

\begin{figure}[htp]
\centering
\includegraphics[width=0.45\textwidth]{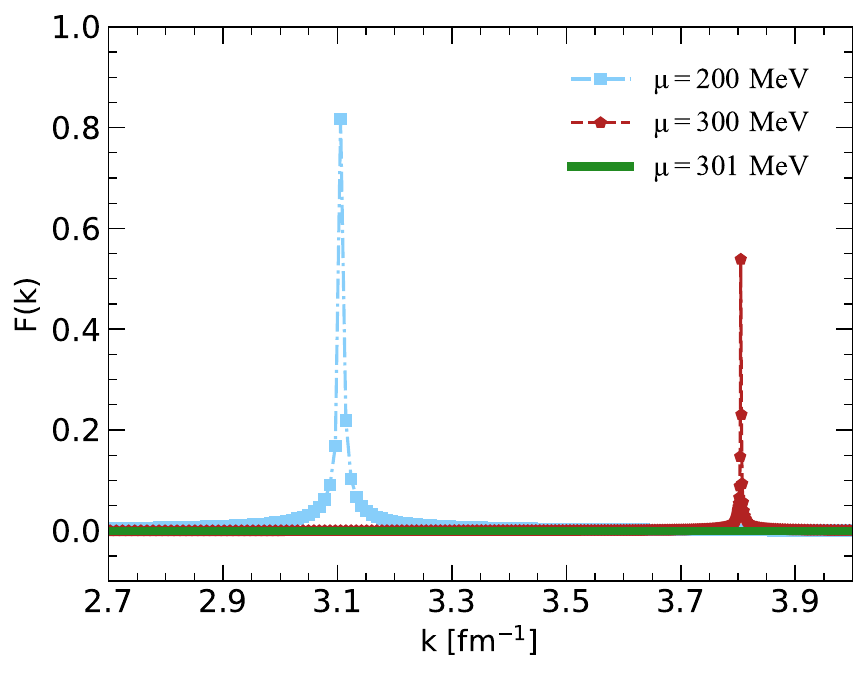}
\caption{$F(k)$ close to a phase transition for the PT potential. The qualitative difference between the form of $F(k)$ on one side of the transition (blue squares and red pentagons) and the other (green solid line) means that the two solutions are well separated ensuring stability.}
\label{fig:trivial_fs}
\end{figure}

Figures~\ref{fig:conv} and \ref{fig:maxf} show ``spikes'' in the two convergence metrics. The monotonicity (or lack thereof) of the convergence metrics depends on the properties of the Jacobian which are in turn prescribed by the potential. In cases where a semi-singular Jacobian arises one has to manually course-correct Newton's method resulting in a ``spike'' in all convergence metrics; for more details see sec.~\ref{sec:smooth}. In the end, this doesn't affect the overall convergence, e.g., the PT potential's convergence exhibits no spikes and requires 21 steps for convergence while the convergence of the \nmlo{2} $(R_0=0.9~\textrm{fm})$ potential exhibits many spikes and requires $24$ steps to converge.

\subsection{Small-gap solutions}
\label{sec:trivial}
Solutions of the gap equations that correspond to small gap functions are typically hard to find when solving Eq.~(\ref{eq:gapeq}) with iteration methods. Typically one encounters this difficulty when dealing with superfluids close to a phase transition to a normal fluid, where the phase's order parameter is very small. A difficulty arises because the trivial solution $\Delta(k)=0$ for all $k$ is always a solution of the gap equations, i.e.,  the BCS wavefunction can also describe a non-superfluid state. Moreover, the gap function is a continuous function of the chemical potential for all $k$, in the region of the phase transition. That is, if $\mu^*$ is the chemical potential where the phase transition happens, then
\begin{align}
    \Delta(k;\mu^\star)=\lim_{\mu\to\mu^\star} \Delta(k;\mu)=0~,\label{eq:ds_lims}
\end{align}
for $k\in\mathbb{R}^+$. (In this section we will make explicit the dependence of $\Delta(k;\mu)$ and $F(k;\mu)$ on $\mu$ by including it as an argument.) In practice, this means that when looking for solutions in the gap equations for chemical potentials in the vicinity of $\mu^\star$ one must be able to tell apart two solutions that are arbitrary close to each other. In other words, $\Delta(k,\mu^\star)$ can be expressed as a small variation of any $\Delta(k,\mu)$ with $|1-\mu^\star/\mu|\ll1$ and the two solutions become almost indistinguishable. The standard solution to this issue is to solve independently for the size of the gap function, $\Delta(k_F)$, and its shape $\Delta(k)/\Delta(k_F)$, as proposed by Khodel, Khodel, and Clark~\cite{khodel:2001}. 
\begin{figure}[htp]
\centering
\includegraphics[width=0.45\textwidth]{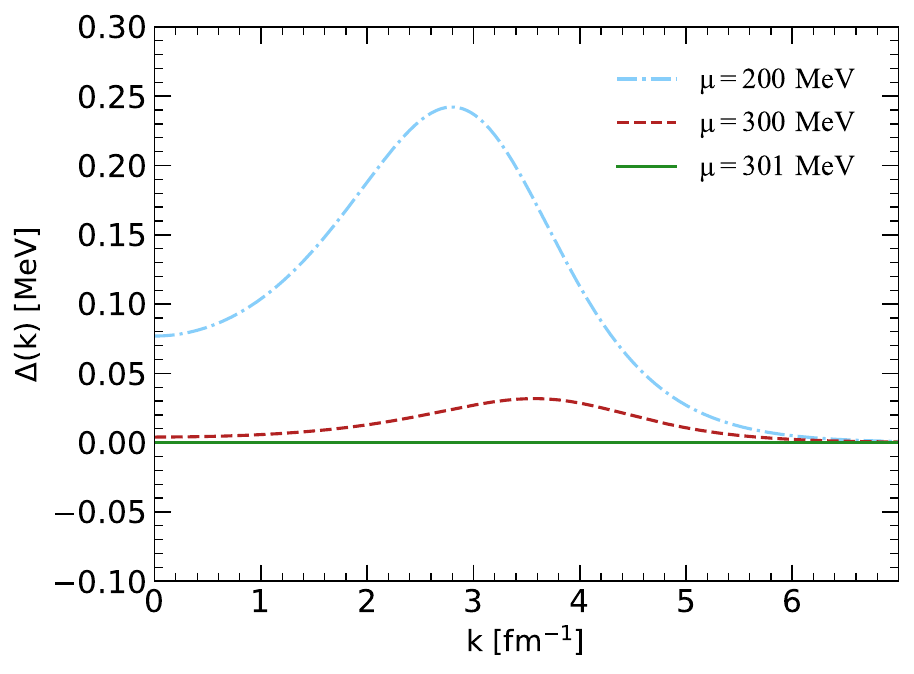}
\caption{Gap functions close to a phase transition for the PT potential, at $\mu^*=301~\textrm{MeV}$ (green line). When approaching a phase transition towards the normal state, the gap function approaches the trivial solution $\Delta(k)=0$ continuously, exemplified by the solutions for $\mu^*=200~\textrm{MeV}$ (blue dashed dotted line) and $\mu^*=300~\textrm{MeV}$ (red dashed line). Close enough to the transition the two solutions come arbitrarily close.}
\label{fig:trivial_ds}
\end{figure}

Recasting of the gap equation in the form of Eq. (\ref{eq:recast}) solves this problem without further manipulation. The quantity that we are solving for, $F(k)=\Delta(k)/E(k)$, is not a continuous function of $\mu$ close to a phase transition. This is because, for $k_0=\sqrt{2m\mu/\hbar^2}$,

\begin{align}
    |F(k_0;\mu)| &= 1~,\quad \textrm{if}~\Delta(k_0;\mu)\neq 0~, \notag \\ &=0~,\quad \textrm{if}~\Delta(k_0;\mu)=0 ~,\label{eq:fs_lims}
\end{align}
which follows directly from Eqs.~(\ref{eq:condensation_amp}) and (\ref{eq:ds_lims}). Then, since $F(K;\mu)$ is a continuous function of $k$ (this, of course, doesn't hold when $k$ is taken on a grid; this is discussed in the next paragraph), there will be at least a finite interval $(k_1,k_2)$ around $k_0$, i.e., $k_0\in(k_1,k_2)$ where
\begin{align}
\lim_{\mu\to\mu^\star}F(k;\mu) \neq F(k;\mu^\star)~, \label{eq:condensation_amp_lim}
\end{align}
for all $k\in(k_1,k_2)$. In other words $F(k,\mu^\star)$ as a function of $k$, cannot be expressed as a small variation of any $F(k,\mu)$ with $\mu\neq\mu^\star$ and so the two solutions remain well separated.

These arguments are based on the limiting behavior of $\Delta(k;\mu)$ and $F(k;\mu)$ and so they hold for the continuum case, when $k\in \mathbb{R}^+$. When using a grid in $k$ these limits have to be taken more carefully, especially the ones in Eq.~(\ref{eq:condensation_amp_lim}). In that case, which is the realistic one, the peak of $F(k;\mu)$ around $k_0\approx \sqrt{2m\mu/\hbar^2}$ might fall in-between grid points and since the peak's width is proportional to $\Delta(k_0)$, close to a phase transition the peak might be missed entirely. In other words, the upper case in Eq.~(\ref{eq:fs_lims}) can be missed entirely if the $k$-grid doesn't contain a point sufficiently close to $k_0$. This issue can be solved by moving the stitch of the first two Legendre grids to $k_0$. Then the increased density of points at the end of a Legendre grid will ensure that the peak is sufficiently mapped. To demonstrate this subtlety, we have included the $k$-grid used for the calculations in Fig.~\ref{fig:trivial_fs} as points. Note that the peaks only reach the value of $F(k;\mu)$ at $k$ closest to the peak's location.

This qualitative difference between the two forms of the gap equation can be seen in Figs.~\ref{fig:trivial_fs} and \ref{fig:trivial_ds} where we plot $F(k;\mu)$ and $\Delta(k;\mu)$, respectively, for the PT potential at $\mu=200~\textrm{MeV}$, $\mu=300~\textrm{MeV}$, and $\mu=301~\textrm{MeV}$. For these calculations we shift the stitch of the first two Legendre grids to $k\approx 3.8~\textrm{fm}^{-1}\approx \sqrt{2m\mu/\hbar^2}$, where $\mu\approx300~\textrm{MeV}$ which brings a dense part of the $k$-grid close to the peak of $F(k;\mu)$; this peak can be determined \textit{a priori} to a good approximation by $\sqrt{2m\mu/\hbar^2}$. To make the connection with the preceding discussion explicit note that when using a PT potential tuned to resemble the neutron-neutron interaction, $\mu\approx301~\textrm{MeV}$ corresponds to a phase transition,i.e., $\Delta(k;\mu=301~\textrm{MeV})=0$ and so $\mu^*=301~\textrm{MeV}$.

From the above discussion it is clear that solving for $F(k;\mu)$ instead of $\Delta(k;\mu)$ is more appropriate especially for small order-parameter solutions, like the ones in Figs.~\ref{fig:trivial_fs} and \ref{fig:trivial_ds}, which is exemplified by those solutions' convergence: at $\mu=300~\textrm{MeV}$, to achieve the same accuracy on a similar $k$-grid by solving for $\Delta(k)$ with a direct iteration method one needs ten times more iterations/steps than the $\sim20$ steps required with recast method. This is a general feature of the new method whose convergence properties don't depend much on the specifics of the system such as the density, the potential, or the magnitude of the superfluid order parameter.

\begin{figure}[htp]
\centering
\includegraphics[width=0.45\textwidth]{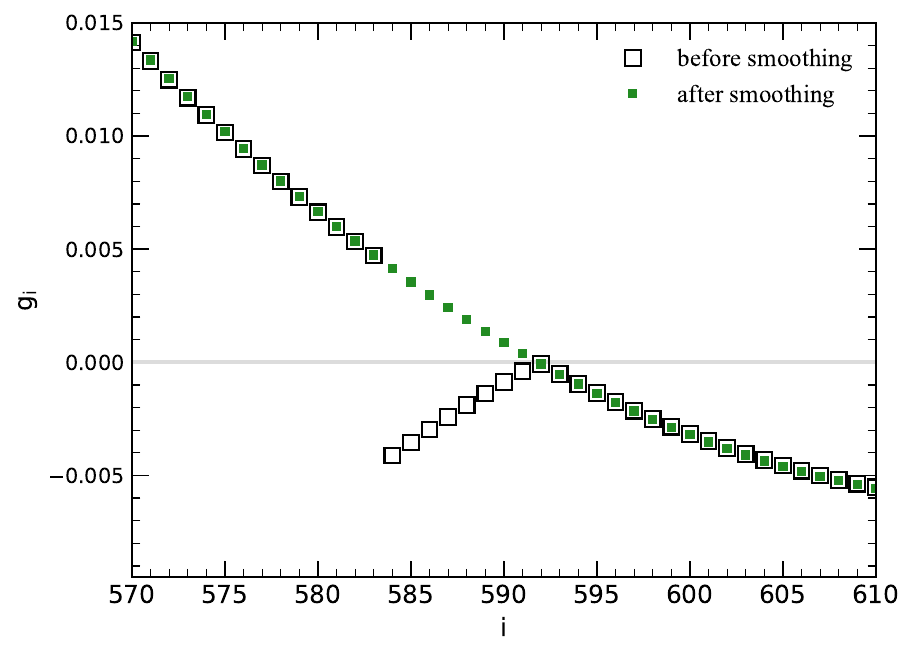}
\caption{ Distributions $g_i$ before and after the ``smoothing'' described in sec.~\ref{sec:smooth}.}
\label{fig:flip}
\end{figure}

\subsection{Spikes in convergence: sign-flips and ``smoothing''}
\label{sec:smooth}
In practice Eq.~(\ref{eq:recast}) can be satisfied by discontinuous distributions $\tilde{g}_i$ that resemble a true solution $\tilde{g}_i$ except for a finite piece that is \textit{almost} related by a sign flip. Indeed, assume a vector $\tilde{g}_i$ and define the partially signed-flipped $\tilde{h}_i$ as
\begin{align}
    \tilde{h}_i &= -\tilde{g}_i~, \quad i\in [i_0,i_1]~, \notag\\
    &= \tilde{g}_i~, \quad \textrm{otherwise}~,
\end{align}
Then
\begin{align}
    f_i(\{\tilde{h}\}) 
    &= f_i(\{\tilde{g}\}) \notag\\
    &+ 4(\tilde{g}^2_i-1)\left(\sum_{j\cancel{\in}[i_0,i_1]} \tilde{\psi}_{ij}\tilde{g}_j \sum_{l\in[i_0,i_1]}\tilde{\psi}_{il}\tilde{g}_l\right)~. \label{eq:flip}
\end{align}
This shows that when using root finding methods that step through the space of $\tilde{g}$ vectors, then in $\tilde{g}_i$ regions where $\tilde{g}_i$ is sufficiently close to 1 or $\psi_{ij}$ is sufficiently close to 0 for a range of $j$ values, they are vulnerable to such sign-flips. In practice this means that often a step with Newton's method will yield a piece-wise discontinuous distribution whose discontinuity can be seen as a result of a piece of it being multiplied by $-1$. We plot one such distribution in Fig.~\ref{fig:flip} where the open squares show the discontinuous distribution generated by a step of Newton's method while the closed symbols show the continuous distribution that one gets after ``smoothing'' the distribution by multiplying the discontinuous piece by $-1$. This is the discontinuity in the green squares of Fig.~\ref{fig:regimes}. When Newton's method seems to be converging to such distributions we manually ``smoothen'' them before continuing as described below.

In principle, in Newton's method, a step that yields a discontinuous distribution should be equally probable with one that corrects the discontinuity: Equation~(\ref{eq:flip}) can also describe the gain/loss in $f_i$ when a discontinuous distribution $\tilde{g}_i$ is corrected. However, converging to a continuous distribution is less likely than a discontinuous one because there are more ways than one to piece-wise break a continuous curve. It's important to note that such discontinuous distributions, like the one in Fig.~\ref{fig:flip}, do not correspond to solutions of the gap equation because they do not satisfy the condition in Eq.~(\ref{eq:cond3}) and indeed they do not satisfy the gap equations. This means that one must course-correct Newton's method when it's attempting to converge to a discontinuous distribution. In practice this is done by enforcing $\tilde{g}_0\tilde{g}_1>0$ and using a forward finite difference at each $i>1$  to estimate the value of $\tilde{g}_{i+1}$, namely $\tilde{\gamma}_{i+1}$. Then if $|\tilde{\gamma}_{i+1}-\tilde{g}_{i+1}| > |-\tilde{g}_{i+1}-g_{i+1}|$, that is $\tilde{g}_{i+1}$ is closer to its approximation when the latter is reflected upon the $x$-axis, we multiply $\tilde{g}_{i+1}$ by $-1$. It might seem problematic that by enforcing $\tilde{g}_0\tilde{g}_1>0$ we are fixing the overall sign of the vector $\tilde{g}_i$, but Eq.~(\ref{eq:recast}) is symmetric under the change of an overall sign in $\tilde{g}_i$. This symmetry is inherited from the original gap equations where for any solution $\Delta(k)$, the function $-\Delta(k)$ is also a solution. In iterative methods, the initial guess for $\Delta(k)$ fixes the ambiguity. We call this process ``smoothing'' and its result is seen in Fig.~\ref{fig:flip} plotted with green diamonds. We find that it's important to apply ``smoothing''' only after some convergence has been achieved and we set that arbitrary threshold in the final regime (see discussion above) to $\delta g < 10^{-4}$, and for every iteration of Newton's method that yields $\delta g < 10^{-4}$, we check for discontinuities and apply the ``smoothing'' if needed. Choosing to course-correct Newton's method in the final regime, after qualitative and quantitative features of $F(k)$ have converged means that we trust that the ``false'' discontinuous solution will be close to the true continuous solution when ``smoothed''. This is corroborated by the relatively short spikes in $\textrm{max}_if_i$ after the ``smoothing'' and the observation that we remain in the convergecne's final regime, seen in Figs~\ref{fig:conv} and \ref{fig:regimes}.  Finally, since ``smoothing'' relies on estimating $\tilde{g}_i$ from $\tilde{g}_{i-1}$ and $\tilde{g}_{i-2}$ via a Taylor expansion it is essential that $F(k)$ doesn't vary rapidly from $k_{i-2}$ through $k_{i}$. At regions of $k$ where large variation is expected a denser grid should be used, e.g., around the narrow peaks discussed in sec.~\ref{sec:trivial}. Ultimately, one doesn't need a prediction from the outset because a small number of such course-corrections are expected to occur and their performance can be easily monitored and assessed, for example by making plots similar to Fig.~\ref{fig:flip}.

``Smoothing'' yields a distribution substantially different form the one that it's applied on, resulting in a spike in any metric of convergence. This can be seen clearly in Fig.~\ref{fig:conv}, where the spike in the \nmlo{3} curve, at the 15th iteration is caused by the exact ``smoothing'' seen in Fig.~(\ref{fig:flip}) (or in Fig.~\ref{fig:regimes}). All other spikes in Fig.~\ref{fig:conv} (or Fig.~\ref{fig:maxf}), correspond to similar ``smoothings''. Finally, as can be seen in Fig.~\ref{fig:flip} and explained with Eq.~(\ref{eq:flip}), a distribution $\tilde{g}_i$ is most likely to ``break'' at $i$ around which $\tilde{g}_i|\ll 1$. This happens either at the tails of the distribution or at its nodes, like in Fig.~\ref{fig:flip}. Occurrences of the former kind do not affect the quality of solution because these regions correspond to high $k$ values where any irregularities are multiplied by the very small value of the potential $V(k',k)$ in Eq.~(\ref{eq:psi_raw}) [or in the gap equation in Eq.~(\ref{eq:gap})]. The discontinuities that appear at the nodes of $\tilde{g}_i$ are the ones that matter and those happen more often in distributions with more nodes which in turn correspond to ``harder'' potentials, i.e., potentials with higher repulsive cores in coordinate space. This is seen in Fig.~(\ref{fig:flip}) where the solution for the PT potential needed no ``smoothing'' (i.e., no spike in $\delta g$), the solution for the \nmlo{3} potential, which has one node at $k\approx1.5~\textrm{fm}^{-1}$ (see Fig.~\ref{fig:gaps}) needed ``smoothing'' once, and the solution to the family of \nmlo{2} potentials, which generally correspond to harder potentials and yield multiple nodes in Fig.~\ref{fig:gaps}, needed ``smoothing'' multiple times before a solution was identified.

\section{Outlook}
\label{sec:conclusions}
Identifying solutions to the BCS theory's gap equations remains a hard problem especially when describing exotic superfluids whose gap functions can have complicated structure. Many approaches have been developed to identify various types of solutions often relying on some \textit{a priori} knowledge of the solution to be identified. In this paper we proposed a new method for solving the gap equations inspired by the work of Refs.~\cite{kumar:1987a,kumar:1987b} on non-linear Hammerstein-type equations. The proposed approach holds the promise of a universal solver for the gap equations agnostic to the properties of the solution to be identified.

In few words, the proposed method recasts the gap equations into a non-linear algebraic equation, seen in Eq.~(\ref{eq:recast}), where one solves for the condensation amplitude $F(k)$, instead of the gap function $\Delta(k)$ on a collocation grid. The resulting set of algebraic equations can be solved by a variety of methods and we have demonstrated its effectiveness using the multidimensional Newton's method for which it outperforms standard approaches; more sophisticated algebraic solvers promise even better performance. 

The good performance of the recast method can be ascribed to a few factors. Solving for the condensation amplitude, i.e., the kernel of the standard gap equations, ensures a rapid convergence which is a general feature of such methods applied to Hammerstein-type equations~\cite{atkinson:1992}. An extra stability is guaranteed for the gap equations close to a phase transition where $\Delta(k)$ is expected to vanish: while the gap function $\Delta(k)$ exhibits a smooth behavior across the transition the condensation amplitude $F(k)$ takes qualitatively different forms on either side of it. Finally, in the recast form of Eq.~(\ref{eq:recast}) the integrals involved in the original form can be evaluated once and stored. This can speed-up the calculations significantly, especially for the cases of hard potentials that yield gap functions with fat tails. 

A final advantage, and one that we did not demonstrate, is that in their recast form, the gap equations can be straightforwardly extended to more exotic cases, (e.g., coupled channels), or appended with various constraints, like the average density. In all these cases, the set of equations in Eq.~(\ref{eq:recast}) would be appended with new $f_i$'s capturing the other channels, or the constraints. Then all $f_i$'s can be treated on equal footing in a way similar to the one presented here without the need for devising nested iteration procedures that often scale poorly and suffer from stability issues.

It's becoming increasingly clear that the rich structure of the nuclear interaction can create exotic phases of nuclear superfluidity defining an unexplored region in the phase space of nuclear many-body systems~\cite{takatsuka:1993,bulgac:2006,ding:2016,palkanoglou:2024,guo:2025,hinohara:2024,lian:2025}. While a proper study of these superfluids requires beyond mean-field techniques, the mean-field description serves as a qualitative guide to the expected physics and so creating the means for reliable such mean-filed descriptions is crucial. The method presented in this paper holds the promise of identifying solutions to the BCS gap equations, the main building block of mean-field descriptions of superfluidity, efficiently and with little \textit{a priori} knowledge of the solution to be identified.

\section{Acknowledgements}
This work was supported by
the Natural Sciences and Engineering Research Council
(NSERC) of Canada and the Canada Foundation for Innovation (CFI). Computational resources were provided by
SHARCNET and NERSC. TRIUMF receives federal funding
via a contribution agreement with the National Research
Council of Canada. Computational resources were provided by SHARCNET and NERSC.

\pagebreak[4]

\appendix

\section{The joint}
\label{app:joint}
As mentioned in sec.~\ref{sec:recast}, mapping the tail of the $k$-grid to infinity via a joint works best for capturing the large-$k$ behavior of $F(k)$. Here we provide the details of using the joint.

Reiterating, the use of a joint corresponds to mapping the tail of the $k$-gird as
\begin{align}
\left[0,\infty\right] &= \left[0,k_0\right]\cup \left[k_0,\infty\right) \notag \\
&\mapsto \left[0,k_0\right]\cup \left[\lambda(k_0),\lambda(\infty)\right] = [t(0),t(\infty)]~,
\end{align} 
where the mapping function $t$ is
\begin{align}
    t(k) &= k~,\quad k\le k_0 \\
        &=\lambda(k)~,\quad  k>k_0
\end{align}
and
\begin{align}
\lambda(k_0),\lambda(\infty) <\infty~.
\end{align}
Continuity of $t$ and its derivative imposes
\begin{align}
    \lambda(k_0) &= k_0~, \label{eq:lamreq1}\\
    \lambda'(k_0) &= 1~,\label{eq:lamreq2}
\end{align}
and we impose the additional requirement $\lambda'(k)>0$ for all $k$. From all possible function $\lambda$ that fit these requirements we choose
\begin{align}
    \lambda(k) &= \lambda_\infty-\frac{\lambda_\infty-\lambda_0}{1+(k-k_0)}~,
\end{align}
where
\begin{align}
    \lambda_0 &= \lambda(k_0)~, \\
    \lambda_\infty &= \lambda(k_\infty)~.
\end{align}
From Eq.~(\ref{eq:lamreq1}), $\lambda_0 = k_0$ and from Eq.~(\ref{eq:lamreq2}) $\lambda_\infty = k_0+1$. Note that $\lambda(k)>k_0~\forall k$, and so there is no $k>k_0$ where $\lambda(k)=k$. This means that mapping the tail of our grid with $\lambda$ won't ``fold'' it on top of existing grid points. We will use the variable $x$ for the mapped grid either before the joint, where $x\equiv k$ or after the joint where $x=\lambda (k)$. Since $\tilde{\lambda}$ is an increasing function of $k$, the inverse exists and we can define the inverse mapping
\begin{align}
t^{-1}(x) &= x ~, \quad x\le x_0   \\
    & = \lambda^{-1}(x)~, \quad x>x_0
\end{align}
with $x_0=k_0$ and 
\begin{align}
    \lambda^{-1}(x) &= k_0 -1 +\frac{1}{k_0+1-x}~.
\end{align}
For clarity we will signify by $N$ the number of grid points in the unmapped grid ($k\le k_0$), enumerated by Latin indices and $M$ the number of grid points in the tail ($k>k_0$) enumerated by Greek indices. We can now define the interpolation on the mapped grid as
\begin{align}
F(k) &= \sum_{a=1}^{N+M} g_j \phi^{(t)}_j(k) \\
\phi_i^{(t)} (k) &=\frac{w_i}{t(k)-t(k_i)} \left[\sum_{j=1}^{N+M}\frac{w_j}{t(k)-t(k_j)}\right]^{-1}
\end{align}
Equivalently, in the mapped space
\begin{align}
    \tilde{F}(x) &= \sum_{i=1}^{N+M} \tilde{g}_i \tilde{\phi}_{i}(x) \\
    \tilde{\phi}(x) &= \frac{w_i}{x-x_i} \left[\sum_{j=1}^{N+M}\frac{w_j}{x-x_j}\right]^{-1}
\end{align}
The relation between the two is
\begin{align}
    \tilde{F}(x) &= F(t^{-1}(x))~, \label{eq:ftilde}\\
    \tilde{\phi}_i(x) &=\phi^{(t)}(t^{-1}(x))~, \label{eq:ltilde}
\end{align}
which allows us to still identify the coefficients $g$ with the values of $F$ on the grid:
\begin{align}
    g_i&=F(k_i)\\
    \tilde{g}_i&=\tilde{F}(x_i)\\
\end{align}
On the transformed grid, the $\psi$ matrix becomes:
\begin{align}
\psi_i(k) &= -\frac{1}{\pi} \int_0^\infty dp p^2 V(k,p)\phi^{(t)}_i(p) \\
&= -\frac{1}{\pi} \int_{0}^{\lambda_\infty} dx \left[t^{-1}(x)\right]' \left[t^{-1}(y)\right]^2 V[k,t^{-1}(x)]\tilde{\phi}_j(x)      
\end{align}
and so
\begin{align}
\tilde{\psi}_i(x) &= \psi_a\left[t^{-1}(x)\right] \notag\\
&= -\frac{1}{\pi} \int_{0}^{\lambda_\infty} dy \left[t^{-1}(y)\right]' \left[t^{-1}(y)\right]^2 V[t^{-1}(x),t^{-1}(y)]\tilde{\phi}^{(t)}_j(y)      
\end{align}
The recast gap equations now become
\begin{align}
\left[\sum_i \tilde{\phi}_i(x) \tilde{g}_i\right]^2 & \left[\xi^2\left[t^{-1}(x)\right]+\left(\sum_i\tilde{\psi}_i(x)\tilde{g}_i\right)^2\right] -\notag \\
&-\left(\sum_i\tilde{\psi}_i(x)\tilde{g}_i\right)^2=0& 
\end{align}
which, when discretized, corresponds to Eq.~(\ref{eq:recast}).


\begin{thebibliography}{99}
\bibitem{dean:2003}
D.~J.~Dean and M.~Hjorth-Jensen, Rev. Mod. Phys. \textbf{75}, 607~(2003).
\bibitem{sedrakian:2019}
A.~Sedrakian and J.W.~Clark, Eur. Phys. J. A \textbf{55}, 167 (2019).
\bibitem{fujimoto:2024} Y. Fujimoto, Phys. Rev. B \textbf{111}, 184510 (2025).
\bibitem{kumamoto:2024} M. Kumamoto and S. Reddy, Phys. Rev. C \textbf{110}, 025804 (2024).
\bibitem{krotscheck:2023}  E. Krotscheck, P. Papakonstantinou, and J. Wang, Astrphys. J. \textbf{955}, 76 (2023).
\bibitem{ding:2016}
D. Ding, A. Rios, H. Dussan, W. H. Dickhoff, S. J. Witte, A. Carbone, and A. Polls, Phys. Rev. C \textbf{94}, 025802 (2016).
\bibitem{burgio:2021} G.F. Burgio, H.J. Schulze, I. Vida{\~n}a, J.B. Wei, Prog. Part. Nucl. Phys. \textbf{120} 103879 (2021).
\bibitem{allard:2025} V. Allard and M. Chamel, Universe \textbf{11}(5), 140 (2025).
\bibitem{tichai:2018} A. Tichai, P. Arthuis, T. Duguet, H. Hergert, V. Som{\' a}, R. Roth, Phys. Lett. B, Physics Letters B \textbf{786}, 195 (2018).
\bibitem{drissi:2024}
M. Drissi, A.~Rios, and C.~Barbieri, Annals of Physics \textbf{469} 169729,  (2024).
\bibitem{broglia:book}
R. A. Broglia and V. Zelevinsky, eds., \textit{Fifty years of nuclear BCS: pairing in finite systems}, World Scientific Publishing, Singapore, (2013).
\bibitem{annett:book}
J.~F.~Annett, \textit{Superconductivity, superfluids and condensates}, Oxford Univ. Press (2004).
\bibitem{almirante:2025} G. Almirante and M. urban, Phys. Rev. Lett. \textbf{135}, 132701 (2025).
\bibitem{gezerlis:2010} 
A.~Gezerlis and J.~Carlson, Phys. Rev. C \textbf{81}, 025803 (2010)
\bibitem{gandolfi:2022} 
S. Gandolfi, G.~Palkanoglou, J.~Carlson, A.~Gezerlis, and
K.~E.~Schmidt, Condensed Matter \textbf{7} (1), 19 (2022).
\bibitem{khodel:1998}
V. V. Khodel, V. A. Khodel and J. W. Clark, Phys. Rev. Lett. \textbf{81}, 3828 (1998).
\bibitem{khodel:2001} V. V. Khodel, V. A. Khodel, and J. W. Clark, Nucl. Phys. A \textbf{679}, 827 (2001).
\bibitem{krotscheck:1972}
E. Krotscheck, Z. Phys. \textbf{251}, 135 (1972).
\bibitem{ramanan:2007}
S. Ramanan, S. Bogner, and R. J. Furnstahl, Nucl. Phys. A \textbf{797}, 81 (2007).
\bibitem{goodman:1972} A. L. Goodman, Nucl. Phys. A \textbf{186}, 475 (1972).
\bibitem{goodman:2001} A. L. Goodman, Phys. Rev. C \textbf{63}, 044325 (2001).
\bibitem{gezerlis:2011} A. Gezerlis, G. F. Bertsch, and Y. L. Luo, Phys. Rev. Lett. \textbf{106}, 252502 (2011).
\bibitem{takatsuka:1993}
T.~Takatsuka, R.~Tamagaki, Prog. of Theor. Phys. Suppl., \textbf{112}, 27 (1993).
\bibitem{guo:2025}
Y.~Guo, T.~Naito, H.~Tajima, H.~Liang Phys. Rev. C \textbf{112}, 024310 (2025).
\bibitem{bulgac:2006}
A.~Bulgac, M.~M.~Forbes, and A.~Schwenk, Phys. Rev. Lett. 97, 020402 (2006).
\bibitem{palkanoglou:2024}
G~Palkanoglou and A.~Gezerlis, Phys. Rev. Lett. \textbf{134}, 032501 (2024).
\bibitem{frauendorf:2014} S. Frauendorf and A. O. Macchiavelli, Prog. Part. Nucl. Phys. \textbf{78}, 24 (2014).
\bibitem{atkinson:1992} K. Atkinson, J. Integral Equ. Appl. \textbf{4} (1) 15–46 (1992).
\bibitem{martin:2016} N. Martin and M. Urban, Phys. Rev. C \textbf{94}, 065801 (2016).
\bibitem{drischler:2017} C. Drischler, T. Kr{\" u}ger, K. Hebeler, and A. Schwenk, Phys. Rev. C \textbf{95}, 024302 (2017).
\bibitem{johnson:1988} D.D.~Johnson, Phys. Rev. B \textbf{38}, 12807 (1988).
\bibitem{kumar:1987a} S. Kumar, IMA J. Numerical Anal. \textbf{7}, 313-326 (1987), 
\bibitem{kumar:1987b} S. Kumar and I. Sloan, Math. Comp. \textbf{48}, 585-593 (1987).
\bibitem{berrut:2004} J.-P. Berrut and L. N. Trefethen, SIAM Rev. \textbf{46}, 501 (2004).
\bibitem{gezerlis:book} A. Gezerlis, \textit{Numerical Methods in Physics with Python, 2nd ed.}, Cambridge University Press; 2023.
\bibitem{pera:2023} J. Pera and J. Boronat, Am. J. Phys. 91, 90 (2023).
\bibitem{gezerlis:2013} A. Gezerlis, I. Tews, E. Epelbaum, S. Gandolfi, K. Hebeler, A. Nogga, and A. Schwenk, Phys. Rev. Lett. 111, 032501 (2013). 
\bibitem{entem:2003} D.R~Entem and R. Machleidt, Phys. Rev. C 68, 041001(R) (2013).
\bibitem{hinohara:2024}
N.~Hinohara, T.~Oishi, and K.~Yoshida, Phys. Rev. C \textbf{109}, 034302 (2024)
\bibitem{lian:2025} X. Lian, C. L. Bai, H. Sagawa, H. Q. Zhang, 	arXiv:2507.18124 [nucl-th].
\end{thebibliography}
\end{document}